\documentstyle[11pt,epsfig]{article}
\title{\bf Bianchi spacetimes in noncommutative phase-space}
\author{ B. Vakili\thanks{email: b-vakili@cc.sbu.ac.ir}, N. Khosravi
\thanks{email: n-khosravi@cc.sbu.ac.ir} and H. R. Sepangi
\thanks{email: hr-sepangi@cc.sbu.ac.ir}
\\ {\small Department of Physics, Shahid Beheshti University, Evin, Tehran 19839, Iran}}
\begin{document}
\maketitle 

\begin{abstract}
The effects of noncommutativity in the phase-space of the
classical and quantum cosmology of Bianchi models are
investigated. Exact solutions in both commutative and
noncommutative cases are presented and compared. Further, the
Noether symmetries of the Bianchi class A spacetimes are studied
in both cases and similarities and differences are discussed.\vspace{3mm}
\end{abstract}
\section{Introduction}
Since 1947 when  noncommutativity between spacetime coordinates
was first introduced by Snyder \cite{1}, a great deal of interest
has been generated in this area of research \cite{2}-\cite{4}.
This interest has been gathering pace in recent years because of
strong motivations in the development of string and M-theories,
\cite{5,6}. However, noncommutative theories may also be justified
in their own right because of the interesting predictions they
have made in particle physics, a few example of which are: the
IR/UV mixing and non-locality \cite{7}, Lorentz violation \cite{8}
and new physics at very short distance scales \cite{9}-\cite{11}.
Noncommutative versions of ordinary quantum \cite{12} and
classical mechanics \cite{13} have also been studied and shown to
be equivalent to their commutative versions if an external
magnetic field is added to the Hamiltonian. The behavior of
relativistic quantum particles (fermions and bosons) in a
noncommutative space is investigated in \cite{14}.

It is a generally accepted practice to introduce noncommutativity
either through the coordinates or fields, which may be called
geometrical or phase-space noncommutativity respectively.
Noncommutative ordinary (quantum) field theories where the
geometry is considered as noncommutative are interesting to study
since they could provide an effective theory bridging the gap
between ordinary quantum field theory and string theory, currently
considered as the most important choice for quantization of
gravity \cite{9}. A different approach to noncommutativity is
through the introduction of noncommutative fields \cite{carmona1},
that is, fields or their conjugate momenta are taken as
noncommuting. These effective theories can address some of the
problems in ordinary quantum field theory, {\it e.g.}
regularization \cite{carmona1} and, predict new phenomenon, such
as Lorentz violation \cite{carmona2}, considered as one of the
general predictions of quantum gravity theories \cite{lorentz}.

Since cosmology can test physics at energies that are much higher
than those which the experiments on earth can achieve, it seems
natural that the effects of quantum gravity could be observed in
this context. Therefore, until a completely satisfactory theory
regarding cosmology  can be afforded by string theory, the study
of the general properties of quantum gravity through cosmological
systems such as the universe seems reasonably promising. There are
several approaches in considering the notion of noncommutativity
in cosmology, which, as was mentioned above could be the best
alternative in the absence of a comprehensive and satisfactory
theory from string theory. These ideas have been studied in
different works, an example of which can be found in
\cite{landi,brand}. There, taking coordinates as noncommuting, it
has been shown that noncommutativity affects the spectrum of
Cosmic Microwave Background \cite{landi,brand}. In \cite{landi},
noncommutative geometry suggests a non-local inflaton field that
changes the gaussianity and isotropy properties of fluctuations.
In this approach the parameter of noncommutativity is a dynamical
quantity. Introduction of noncommutativity in \cite{brand} is
based on the stringy spacetime uncertainty principle (SSUR), see
also \cite{b12}. This approach results in the appearance of a
critical time for each mode at which SSUR is saturated, and which
is taken to be the time the mode is generated. In this case, in
contrast to \cite{landi}, non-locality appears  in time. In both
the above approaches the spacetime is taken to be the isotropic
FRW spacetime. We follow  a phase-space approach to
noncommutativity in cosmology, like that presented in
\cite{carmona1} which is different from what has been studied in
\cite{landi,brand} and start with Bianchi models as anisotropic
spacetimes rather than an isotropic universe.

In cosmological systems, since the scale factors, matter fields
and their conjugate momenta play the role of dynamical variables
of the system, introduction of noncommutativity by adopting the
approach discussed above is particularly relevant. The resulting
noncommutative classical and quantum cosmology of such models have
been studied in different works \cite{15}. These and similar works
have opened a new window through which some of problems related to
cosmology can be looked at and, hopefully, resolved. For example,
an investigation of the cosmological constant problem can be found
in \cite{16}. In \cite{17} the same problem is carried over to the
Kaluza-Klein cosmology. The problem of compactification and
stabilization of the extra dimensions in multidimensional
cosmology may also be addressed using noncommutative ideas
\cite{18}.

In this paper we deal with noncommutativity in  Bianchi class A
models. The classical and quantum solutions of Bianchi models have
been studied by many authors, see for example \cite{19}-\cite{22}.
Since the Bianchi models have different scale factors in different
directions, they are suitable candidates for studying
noncommutative cosmology. Here, our aim is to introduce
noncommutative scale factors in Bianchi spacetimes and compare and
contrast their solutions to that of the commutative case at both
the classical and quantum levels. We study the effects of
noncommutativity on the underlying symmetries of these models
\cite{23}-\cite{25} and, show that noncommutativity manifests
itself in changing the number of such symmetries. It should be
emphasized that when we speak of noncommutativity in this work, we
mean noncommutativity in the fields (scale factors) and not the
coordinates, that is to say that we study noncommutativity within
the context of phase-space only.
\section{The Model}
Let us make a quick review of some of the  important results in the
Bianchi class A models and obtain their Lagrangian and Hamiltonian
in the ADM decomposition, for more details see \cite{19}. The
Bianchi models are the most general homogeneous cosmological
solutions of the Einstein field equations which admit a
3-dimensional isometry group, \textit{i.e.} their spatially
homogeneous sections are invariant under the action of a
3-dimensional Lie group. In the Misner \cite{26} notation, the
metric of the Bianchi models can be written as
\begin{equation}\label{A}
ds^2=-N^2(t)dt^2+e^{2u(t)}e^{2\beta_{ij}(t)}\omega^i\otimes
\omega^j,
\end{equation}
where $N(t)$ is the lapse function, $e^{2u(t)}$ is the scale factor
of the universe and $\beta_{ij}$ determine the anisotropic
parameters $v(t)$ and $w(t)$ as follows
\begin{equation}\label{B}
\beta_{ij}=\mbox{diag}\left(v+\sqrt{3}w,v-\sqrt{3}w,-2v\right).
\end{equation}
Also, in metric (\ref{A}), the 1-forms $\omega^i$ represent the
invariant 1-forms of the corresponding isometry group and satisfy
the following Lie algebra
\begin{equation}\label{C}
d\omega^i=-\frac{1}{2}C^i_{jk}\omega^j \wedge\omega^k,\end{equation}
where $C^i_{jk}$ are the structure constants. Indeed, the Bianchi
models are grouped by their structure constants into classes A and
B. Because of the difficultly in formulating the class B Bianchi
models in the context of the ADM decomposition and canonical
quantization \cite{27}, it is usually the case that one confines
attention to the class A models where the structure constants obey
the relation $C^i_{ji}=0$.

The Einstein-Hilbert action is given by (we work in units where
$c=\hbar=16\pi G=1$)
\begin{equation}\label{D}
{\cal S}=\int d^4 x\sqrt{-g}({\cal R}-\Lambda),\end{equation}
where $g$ is the determinant of the metric, ${\cal R}$ is the
scalar curvature of the spacetime metric (\ref{A}) and $\Lambda$
is the cosmological constant. In terms of the ADM variables,
action (\ref{D}) can be written as \cite{28}
\begin{equation}\label{E}
{\cal S}=\int dt d^3x{\cal L}=\int dt d^3x
N\sqrt{h}\left(K_{ij}K^{ij}-K^2+R-\Lambda\right),\end{equation}
where $K_{ij}$ are the components of extrinsic curvature (second
fundamental form) which represent how much the spatial space
$h_{ij}$ is curved in the way it sits in the spacetime manifold.
Also, $h$ and $R$ are the determinant and scalar curvature of the
spatial geometry $h_{ij}$ respectively, and $K$ represents the
trace of $K_{ij}$. The extrinsic curvature is given by
\begin{equation}\label{F}
K_{ij}=\frac{1}{2N}\left(N_{i|j}+N_{j|i}-\frac{\partial
h_{ij}}{\partial t}\right),\end{equation}where $N_{i|j}$ represents
the covariant derivative with respect to $h_{ij}$. Using (\ref{A})
and (\ref{B}) we obtain the non-vanishing components of the
extrinsic curvature and its trace as follows
\begin{eqnarray}\label{G}
K_{11}&=&-\frac{1}{N}(\dot{u}+\dot{v}+\sqrt{3}\dot{w})e^{2(u+v+\sqrt{3}w)},\nonumber \\
K_{22}&=&-\frac{1}{N}(\dot{u}+\dot{v}-\sqrt{3}\dot{w})e^{2(u+v-\sqrt{3}w)},\nonumber\\
\\
K_{33}&=& -\frac{1}{N}(\dot{u}-2\dot{v})e^{2(u-2v)},\nonumber\\
K&=&-3\frac{\dot{u}}{N}, \nonumber
\end{eqnarray}
where a dot represents differentiation with respect to $t$. The
scalar curvature $R$ of a  spatial hypersurface is a function of $v$
and $w$ and can be written in terms of the structure constants as
\cite{21}
\begin{equation}\label{H}
R=C^i_{jk}C^l_{mn}h_{il}h^{km}h^{jn}+2C^i_{jk}C^k_{li}h^{jl}.
\end{equation}

The Lagrangian for the Bianchi class A models may now be written by
substituting the above results into action (\ref{E}), giving
\begin{equation}\label{I}
{\cal
L}=\frac{6e^{3u}}{N}\left(-\dot{u}^2+\dot{v}^2+\dot{w}^2\right)+Ne^{3u}\left(R-\Lambda
\right).
\end{equation}
The momenta conjugate to the dynamical variables are given by
\begin{equation}\label{J}
p_u=\frac{\partial {\cal L}}{\partial
\dot{u}}=-\frac{12}{N}\dot{u}e^{3u},\hspace{.5cm}p_v=\frac{\partial
{\cal L}}{\partial
\dot{v}}=\frac{12}{N}\dot{v}e^{3u},\hspace{.5cm}p_w=\frac{\partial
{\cal L}}{\partial \dot{w}}=\frac{12}{N}\dot{w}e^{3u},
\end{equation}
leading to the following Hamiltonian
\begin{equation}\label{K}
{\cal
H}=\frac{1}{24}Ne^{-3u}\left(-p_u^2+p_v^2+p_w^2\right)-Ne^{3u}\left(R-\Lambda\right).
\end{equation}
The preliminary set-up for writing the dynamical equations at both
the classical and quantum levels is now complete. In what follows,
we will study these equations in commutative and noncommutative
cases.
\section{Classical cosmology}
\subsection{Commutative case}
The classical and quantum solutions of the Bianchi models are
studied in many works \cite{19}-\cite{22}. Since our aim here is to
compare these solutions to the solutions of the noncommutative
model, in the following two sections we consider only the simplest
Bianchi class A model, namely type I. The structure constants of the
Bianchi type I are all zero, that is $C^i_{jk}=0$. It then follows
from equation (\ref{H}) that  $R=0$. Thus, with the choice of the
harmonic time gauge $N=e^{3u}$ \cite{29}, the Hamiltonian can be
write as
\begin{equation}\label{L}
{\cal H}=\frac{1}{24}\left(-p_u^2+p_v^2+p_w^2\right)+\Lambda e^{6u}.
\end{equation}
The Poisson brackets for the classical phase-space variables are
\begin{equation}\label{M}
\{x_i,x_j\}=0,\hspace{.5cm}\{p_i,p_j\}=0,\hspace{.5cm}\{x_i,p_j\}=\delta_{ij},
\end{equation}
where $x_i(i=1,2,3)=u,v,w$ and $p_i(i=1,2,3)=p_u,p_v,p_w$.
Therefore, the equations of motion become
\begin{equation}\label{O}
\dot{u}=\{u,{\cal
H}\}=-\frac{1}{12}p_u,\hspace{.5cm}\dot{p_u}=\{p_u,{\cal
H}\}=-6\Lambda e^{6u},
\end{equation}
\begin{equation}\label{P}
\dot{v}=\{v,{\cal
H}\}=\frac{1}{12}p_v,\hspace{.5cm}\dot{p_v}=\{p_v,{\cal H}\}=0,
\end{equation}
\begin{equation}\label{Q}
\dot{w}=\{w,{\cal
H}\}=\frac{1}{12}p_w,\hspace{.5cm}\dot{p_w}=\{p_w,{\cal H}\}=0.
\end{equation}
Equations (\ref{P}) and (\ref{Q}) can be immediately integrated to
yield
\begin{equation}\label{R}
p_v=p_{0v},\hspace{.5cm}v(t)=\frac{1}{12}p_{0v}t+v_0,
\end{equation}
\begin{equation}\label{S}
p_w=p_{0w},\hspace{.5cm}w(t)=\frac{1}{12}p_{0w}t+w_0,
\end{equation}
where $v_0$, $w_0$, $p_{0v}$ and $p_{0w}$ are integrating constants.
To integrate equations (\ref{O}) we note that the first integral of
this system gives
\begin{equation}\label{T}
\dot{p_u}=4A^2-\frac{1}{4}p_u^2,
\end{equation}
where $A$ is a constant. Now, integration of this equation depends
on the sign of the cosmological constant $\Lambda$. If $\Lambda>0$,
then from the second equation of (\ref{O}) we have $\dot{p_u}<0$,
and therefore equations (\ref{T}) and (\ref{O}) result in
\begin{equation}\label{U}
p_u(t)=4A\coth A(t+t_0),
\end{equation}
\begin{equation}\label{V}
u(t)=\frac{1}{6}\ln
\frac{2A^2}{3\Lambda}\left[\coth^2A(t+t_0)-1\right],
\end{equation}
where $t_0$ is another constant of integration. In the case of a
negative cosmological constant, $\Lambda<0$, we have $\dot{p_u}>0$,
and integration of equations (\ref{T}) and (\ref{O}) leads to
\begin{equation}\label{W}
p_u(t)=4A\tanh A(t+t_0),
\end{equation}
\begin{equation}\label{X}
u(t)=\frac{1}{6}\ln
\frac{2A^2}{-3\Lambda}\left[1-\tanh^2A(t+t_0)\right].
\end{equation}
Finally for a universe with zero cosmological constant, $\Lambda=0$,
we have $\dot{p_u}=0$ and the equation of motion for $u$ is similar
to that of $v$ and $w$
\begin{equation}\label{Y}
p_u=p_{0u},\hspace{.5cm}u(t)=-\frac{1}{12}p_{0u}t+u_0.
\end{equation}
For a non-zero cosmological constant $\Lambda$ these solutions show
a singularity at $u\rightarrow -\infty$ ($e^{2u}\rightarrow 0$) as
$t\rightarrow \infty$, \textit{i.e.} the universe evolves to a
vanishing spatial size. In the case of a zero cosmological constant
this behavior depends on the sign of $p_{0u}$. For positive $p_{0u}$
we have again a similar singularity as before, but for a negative
value of $p_{0u}$, the scale factor goes to infinity as
$t\rightarrow \infty$. While independent of the cosmological
constant's value, the anisotropic parameters increase to infinity.
\subsection{Noncommutative case}
Let us now concentrate on the noncommutativity concepts in classical
cosmology. Noncommutativity in classical physics \cite{13} is
described by a deformed product, also known as the Moyal product law
between two arbitrary functions of position and momenta as
\begin{equation}\label{Z}
(f*_{\alpha}g)(x)=\exp\left[\frac{1}{2}\alpha^{ab}\partial^{(1)}_a
\partial^{(2)}_b\right]f(x_1)g(x_2)|_{x_1=x_2=x},\end{equation}such
that
\begin{equation}\label{AA}
\alpha_{ab}=\left(%
\begin{array}{cc}
\theta_{ij} & \delta_{ij}+\sigma_{ij} \\
-\delta_{ij}-\sigma_{ij} & \beta_{ij} \\
\end{array}%
\right),\end{equation} where the $N\times N$ matrices $\theta$ and
$\beta$ are assumed to be antisymmetric with $2N$ being the
dimension of the classical phase-space and represent the
noncommutativity in coordinates and momenta respectively. With
this product law, the deformed Poisson brackets can be written as
\begin{equation}\label{AB}
\{f,g\}_\alpha=f*_\alpha g-g*_\alpha f.\end{equation} A simple
calculation shows that
\begin{equation}\label{AC}
\{x_i,x_j\}_\alpha=\theta_{ij},\hspace{.5cm}\{x_i,p_j\}_\alpha=\delta_{ij}+\sigma_{ij},\hspace{.5cm}\{p_i,p_j\}_\alpha=\beta_{ij}.
\end{equation}
Now, consider the following transformations on the classical
phase-space
\begin{equation}\label{AD}
x'_i=x_i-\frac{1}{2}\theta_{ij}p^j,\hspace{.5cm}p'_i=p_i+\frac{1}{2}\beta_{ij}x^j.
\end{equation}
It can easily be checked that if $(x_i,p_j)$ obey the usual Poisson
algebra (\ref{M}), then
\begin{equation}\label{AE}
\{x'_i,x'_j\}=\theta_{ij},\hspace{.5cm}\{x'_i,p'_j\}=\delta_{ij}+\sigma_{ij},\hspace{.5cm}\{p'_i,p'_j\}=\beta_{ij},
\end{equation}
where
$\sigma_{ij}=-\frac{1}{8}\left(\theta_i^k\beta_{kj}+\beta_i^k\theta_{kj}\right)$.
These commutative relations are the same as (\ref{AC}).
Consequently, for introducing noncommutativity, it is more
convenient to work with Poisson brackets (\ref{AE}) than
$\alpha$-star deformed Poisson brackets (\ref{AC}). It is important
to note that the relations represented by equations (\ref{AC}) are
defined in the spirit of the Moyal product given above. However, in
the relations defined by (\ref{AE}), the variables $(x_i,p_j)$ obey
the usual Poisson bracket relations so that the two sets of deformed
and ordinary Poisson brackets represented by relations (\ref{AC})
and (\ref{AE}) should be considered as distinct.

In this work we consider a noncommutative phase-space in which
$\beta_{ij}=0$ and so $\sigma_{ij}=0$, \textit{i.e.} the Poisson
brackets of the phase space variables are as follows
\begin{eqnarray}\label{AF}
\{u_{nc},v_{nc}\}=\theta_1,\hspace{.5cm}\{v_{nc},w_{nc}\}=\theta_2,\hspace{.5cm}\{w_{nc},u_{nc}\}=\theta_3,\nonumber\\
\\
\{x_{inc},p_{jnc}\}=\delta_{ij},\{p_{inc},p_{jnc}\}=0.\nonumber
\end{eqnarray}
With the noncommutative phase-space defined above, we consider the
Hamiltonian of the noncommutative model as having the same
functional form as equation (\ref{L}), but in which the dynamical
variables satisfy the above deformed Poisson brackets, that is
\begin{equation}\label{AG}
{\cal
H}_{nc}=\frac{1}{24}\left(-p_{u_{nc}}^2+p_{v_{nc}}^2+p_{w_{nc}}^2\right)+\Lambda
e^{6u_{nc}}.
\end{equation}
Therefore, the equations of motion read
\begin{equation}\label{AH}
\dot{u}_{nc}=\{u_{nc},{\cal
H}_{nc}\}=-\frac{1}{12}p_{u_{nc}},\hspace{.5cm}\dot{p}_{u_{nc}}=\{p_{u_{nc}},{\cal
H}_{nc}\}=-6\Lambda e^{6u_{nc}},
\end{equation}
\begin{equation}\label{AL}
\dot{v}_{nc}=\{v_{nc},{\cal
H}_{nc}\}=\frac{1}{12}p_{v_{nc}}-6\theta_1 \Lambda
e^{6u_{nc}},\hspace{.5cm}\dot{p}_{v_{nc}}=\{p_{v_{nc}},{\cal
H}_{nc}\}=0,
\end{equation}
\begin{equation}\label{AI}
\dot{w}_{nc}=\{w_{nc},{\cal
H}_{nc}\}=\frac{1}{12}p_{w_{nc}}+6\theta_3 \Lambda
e^{6u_{nc}},\hspace{.5cm}\dot{p}_{w_{nc}}=\{p_{w_{nc}},{\cal
H}_{nc}\}=0.\end{equation} Equations (\ref{AH}) are similar to
equations (\ref{O}) in the commutative case. Their solutions are
therefore as follows
\begin{eqnarray}\label{AJ}
u_{nc}(t)&=&\frac{1}{6}\ln \frac{2A^2}{3\Lambda}\left[\coth^2A(t+t_0)-1\right],\nonumber \\
\\
p_{u_{nc}}(t)&=&4A\coth A(t+t_0),\nonumber
\end{eqnarray}
for positive cosmological constant, and
\begin{eqnarray}\label{AL1}
u_{nc}(t)&=&\frac{1}{6}\ln \frac{2A^2}{-3\Lambda}\left[1-\tanh^2A(t+t_0)\right],\nonumber \\
\\
p_{u_{nc}}(t)&=&4A\tanh A(t+t_0), \nonumber
\end{eqnarray}
in the case of a negative cosmological constant. Also for a zero
cosmological constant the solutions can be written as
\begin{equation}\label{AM}
u_{nc}(t)=-\frac{1}{12}p_{0u}t+u_0,\hspace{.5cm}p_{u_{nc}}=p_{0u}.
\end{equation}
Substituting the above results in equations (\ref{AL}) and
(\ref{AI}) yields the following equations for $v_{nc}(t)$ and
$w_{nc}(t)$
\begin{eqnarray}\label{AN}
v_{nc}(t)&=&\frac{1}{12}p_{0v}t+4\theta_1 A\coth A(t+t_0),\hspace{.5cm}p_{v_{nc}}=p_{0v},\nonumber \\
\\
w_{nc}(t)&=&\frac{1}{12}p_{0w}t-4\theta_3 A\coth
A(t+t_0),\hspace{5mm}p_{w_{nc}}=p_{0w}, \nonumber
\end{eqnarray}
if $\Lambda>0$, and
\begin{eqnarray}\label{AO}
v_{nc}(t)=\frac{1}{12}p_{0v}t+4\theta_1 A\tanh A(t+t_0),\hspace{.5cm}p_{v_{nc}}=p_{0v},\nonumber \\
\\
w_{nc}(t)=\frac{1}{12}p_{0w}t-4\theta_3 A\tanh
A(t+t_0),\hspace{.5cm}p_{w_{nc}}=p_{0w}, \nonumber
\end{eqnarray}
if $\Lambda<0$. Finally if $\Lambda=0$, the dynamical equations for
$v(t)$ and $w(t)$ are again similar to the commutative case with
solutions
\begin{eqnarray}\label{AP}
v_{nc}(t)=\frac{1}{12}p_{0v}t+v_0,\hspace{.5cm}p_{v_{nc}}=p_{0v},\nonumber \\
\\
w_{nc}(t)=\frac{1}{12}p_{0w}t+w_0,\hspace{.5cm}p_{w_{nc}}=p_{0w}.
\nonumber\
\end{eqnarray}
As mentioned before, instead of dealing with the noncommutative
variables we can construct, with the help of the transformations
(\ref{AD}),  a set of commutative dynamical variables
$\{u,v,w,p_u,p_v,p_w\}$ obeying the usual Poisson brackets (\ref{M})
which, for the problem at hand read
\begin{eqnarray}\label{AR}
p_{u_{nc}}&=&p_u,\hspace{.5cm}p_{v_{nc}}=p_v,\hspace{.5cm}p_{w_{nc}}=p_w,\nonumber\\
u_{nc}&=&u-\frac{1}{2}\theta_1 p_v+\frac{1}{2}\theta_3p_w,\nonumber\\
\\
v_{nc}&=&v+\frac{1}{2}\theta_1 p_u-\frac{1}{2}\theta_2 p_w,\nonumber\\
w_{nc}&=&w-\frac{1}{2}\theta_3 p_u+\frac{1}{2}\theta_2 p_v.\nonumber
\end{eqnarray}
In terms of these commutative variables the Hamiltonian takes the
form
\begin{equation}\label{AS}
{\cal H}=\frac{1}{24}\left(-p_u^2+p_v^2+p_w^2\right)+\Lambda
e^{6(u-\frac{1}{2}\theta_1p_v+\frac{1}{2}\theta_3p_w)}.
\end{equation}Therefore, we have the following equations of motion
\begin{equation}\label{AT}
\dot{u}=\{u,{\cal
H}\}=-\frac{1}{12}p_u,\hspace{.5cm}\dot{p}_u=\{p_u,{\cal
H}\}=-6\Lambda
e^{6(u-\frac{1}{2}\theta_1p_v+\frac{1}{2}\theta_3p_w)},
\end{equation}
\begin{equation}\label{AU}
\dot{v}=\{v,{\cal H}\}=\frac{1}{12}p_v-3\Lambda \theta_1
e^{6(u-\frac{1}{2}\theta_1p_v+\frac{1}{2}\theta_3p_w)}
,\hspace{.5cm}\dot{p}_v=\{p_v,{\cal H}\}=0,
\end{equation}
\begin{equation}\label{AV}
\dot{w}=\{w,{\cal H}\}=\frac{1}{12}p_w+3\Lambda \theta_3
e^{6(u-\frac{1}{2}\theta_1p_v+\frac{1}{2}\theta_3p_w)}
,\hspace{.5cm}\dot{p}_w=\{p_w,{\cal H}\}=0.
\end{equation}
The solutions of the above equations can be straightforwardly
obtained in the same manner as that of the system
(\ref{O})-(\ref{Q}). It is easy to check that the action of
transformations (\ref{AR}) on  the solutions of system
(\ref{AT})-(\ref{AV}) recovers solutions (\ref{AJ})-(\ref{AP}).

The differences between classical commutative and noncommutative
cosmologies are now notable. It is clear from solutions in these two
cases that both models have a similar singularity at $t\rightarrow +
\infty$. Thus, noncommutativity cannot remove this singular behavior
from the commutative solutions. The other feature of these solutions
is related to their isotropic behavior. With the choice of special
initial conditions $v(0)=p_v(0)=0$ and $ w(0)=p_w(0)=0$ for the case
$\Lambda\neq0$, the commutative classical cosmology predicts an
isotropic universe, while applying the same initial conditions on
the noncommutative solutions yields an anisotropic cosmology.
\section{Quantum cosmology}
\subsection{Commutative case}
Now, let us quantize the model described above. The quantum
cosmology of Bianchi models is well studied, see
\cite{19}-\cite{22}. Here, for comparison purposes and studying the
noncommutativity effects on the solutions, we first discuss the
commutative quantum cosmology of our model. For this purpose, we
quantize the dynamical variables of the model with the use of the
canonical quantization procedure that leads to the Wheeler-DeWitt
(WD) equation ${\cal H}\Psi=0$. Here, ${\cal H}$ is the operator
form of the Hamiltonian given by (\ref{L}) and $\Psi$ is the wave
function of the universe, a function of spatial geometry and matter
fields, if they exist. With the replacement $p_u\rightarrow
-i\frac{\partial}{\partial u}$ and similarly for $p_v$ and $p_w$ in
(\ref{L}) the WD equation reads
\begin{equation}\label{AX}
\left[\frac{\partial^2}{\partial u^2}-\frac{\partial^2}{\partial
v^2}-\frac{\partial^2}{\partial w^2}+24\Lambda
e^{6u}\right]\Psi(u,v,w)=0.
\end{equation}
The solutions of the above differential equation are separable and
may be written in the form $\Psi(u,v,w)=U(u)V(v)W(w)$, leading to
\begin{equation}\label{AY}
\frac{1}{V}\frac{d^2V}{dv^2}=\pm
\eta^2,\hspace{.5cm}\frac{1}{W}\frac{d^2W}{dw^2}=\pm \kappa^2,
\end{equation}
\begin{equation}\label{AZ}
\frac{d^2U}{du^2}+\left(24\Lambda e^{6u}\pm 9\nu^2\right)U=0,
\end{equation}
where $\eta$ and $\kappa$ are separation constants and
$9\nu^2=\eta^2+\kappa^2$. Equations (\ref{AY}) have simple solutions
in the form of exponential functions $e^{\pm i\eta v}$ (or $e^{\pm
\eta v}$) and $e^{\pm i\kappa w}$ (or $e^{\pm \kappa w}$). Also the
solutions of equation (\ref{AZ}) can be written in terms of the
Bessel functions as
$$U(u)=J_{\nu}\left(2\sqrt{\frac{2\Lambda}{3}}e^{3u}\right),$$ for
a positive cosmological constant and
$$U(u)=K_{i\nu}\left(2\sqrt{\frac{2|\Lambda|}{3}}e^{3u}\right),$$ for a
negative cosmological constant. Thus, the eigenfunctions of the WD
equation can be written as
\begin{equation}\label{AAA}
\Psi_{\nu}(u,v,w)=e^{-(\eta v+\kappa
w)}J_{\nu}\left(2\sqrt{\frac{2\Lambda}{3}}e^{3u}\right),\hspace{.5cm}\Lambda>0,
\end{equation}
\begin{equation}\label{AAB}
\Psi_{\nu}(u,v,w)=e^{i(\eta v+\kappa
w)}K_{i\nu}\left(2\sqrt{\frac{2|\Lambda|}{3}}e^{3u}\right),\hspace{.5cm}\Lambda<0,
\end{equation}
where for having well defined functions we use the separation
constants (\ref{AY}) with plus sign when $\Lambda >0$ and minus sign
in the case $\Lambda <0$. We may now write the general solutions to
the WD equations as a superposition of the eigenfunctions
\begin{equation}\label{AAC}
\Psi(u,v,w)=\int_{-\infty}^{+\infty}C(\nu)\Psi_{\nu}(u,v,w)d\nu,
\end{equation}
where $C(\nu)$ can be chosen as a shifted Gaussian weight function
$e^{-a(\nu-b)^2}$ \cite{15}.
\subsection{Noncommutative case}
To study noncommutativity at the quantum level we follow the same
procedure as before, namely the canonical transition from classical
to quantum mechanics by replacing the Poisson brackets with the
corresponding Dirac commutators as $\{\hspace{2mm}\}\rightarrow
-i[\hspace{2mm}]$. Thus, the commutation relations between our
dynamical variables should be modified as follows
\begin{equation}\label{AAD}
[u_{nc},v_{nc}]=i\theta_1,\hspace{.5cm}[v_{nc},w_{nc}]=i\theta_2,
\hspace{.5cm}[w_{nc},u_{nc}]=i\theta_3,\hspace{.5cm}
[u_{nc},p_u]=[v_{nc},p_v]=[w_{nc},p_w]=i.
\end{equation}
The corresponding WD equation can be obtained by the modification of
the operator product in (\ref{AX}) with the Moyal deformed product
\cite{15}
\begin{equation}\label{AAE}
\left[-p_u^2+p_v^2+p_w^2+24\Lambda e^{6u}\right]*\Psi(u,v,w)=0.
\end{equation}
Using the definition of the Moyal product (\ref{Z}), it may be shown
that \cite{15}
\begin{equation}\label{AAF}
f(u,v,w)*\Psi(u,v,w)=f(u_{nc},v_{nc},w_{nc})\Psi(u,v,w),
\end{equation}
where the relations between the noncommutative variables
$u_{nc},v_{nc},w_{nc}$ and commutative variables $u,v,w$ are given
by (\ref{AR}). Therefore, the noncommutative version of the WD
equation can be written as
\begin{equation}\label{AAG}
\left[\frac{\partial ^2}{\partial u^2}-\frac{\partial^2}{\partial
v^2}-\frac{\partial^2}{\partial w^2}+24\Lambda
e^{6(u-\frac{1}{2}\theta_1 p_v+\frac{1}{2}\theta_3
p_w)}\right]\Psi(u,v,w)=0.
\end{equation}
We again separate the solutions into the form $\Psi(u,v,w)=e^{i(\eta
v+\kappa w)}U(u)$. Noting that
\begin{eqnarray}\label{AAH}
 e^{6(u-\frac{1}{2}\theta_1 p_v+\frac{1}{2}\theta_3 p_w)}\Psi(u,v,w)&=&e^{6u}\Psi(u,v+3i\theta_1,w-3i\theta_3)\nonumber\\
 &=&e^{6u}U(u)e^{i\eta(v+3i\theta_1)}e^{i\kappa(w-3i\theta_3)} \nonumber \\
 &=&e^{6u}U(u)e^{i(\eta v+\kappa w)}e^{-3(\eta \theta_1-\kappa \theta_3)}\nonumber \\
 &=&e^{6u}e^{-3(\eta \theta_1-\kappa \theta_3)}\Psi(u,v,w),
\end{eqnarray}
equation (\ref{AAG}) takes the form
\begin{equation}\label{AAI}
\left[\frac{\partial ^2}{\partial u^2}-\frac{\partial^2}{\partial
v^2}-\frac{\partial^2}{\partial w^2}+24\Lambda e^{6u}e^{-3(\eta
\theta_1-\kappa \theta_3)}\right]\Psi(u,v,w)=0.
\end{equation}
Following the solutions to equation (\ref{AX}), we can find the
general solutions for the above equation as
\begin{equation}\label{AAJ}
\Psi(u,v,w)=\int_{-\infty}^{+\infty}A(\nu)\Psi_{\nu}(u,v,w)d\nu,
\end{equation}
where again, $A(\nu)$ is chosen as a shifted Gaussian function and
$\Psi_{\nu}(u,v,w)$ are the eigenfunctions of equation (\ref{AAI}),
that is
\begin{equation}\label{AAL}
\Psi_{\nu}(u,v,w)=e^{-(\eta v+\kappa
w)}J_{\nu}\left(2\sqrt{\frac{2\Lambda}{3}}e^{3(u-\frac{1}{2}\eta
\theta_1+\frac{1}{2}\kappa \theta_3)}\right),\hspace{.5cm}\Lambda>0,
\end{equation}
\begin{equation}\label{AAM}
\Psi_{\nu}(u,v,w)=e^{i(\eta v+\kappa
w)}K_{i\nu}\left(2\sqrt{\frac{2|\Lambda|}{3}}e^{3(u-\frac{1}{2}\eta
\theta_1+\frac{1}{2}\kappa \theta_3)}\right),\hspace{.5cm}\Lambda<0.
\end{equation}
where again $9\nu^2=\eta^2+\kappa^2$. Figure 1 shows the square of
wave functions of the commutative and noncommutative universes. The
comparison of the figures show that the noncommutativity causes a
shift in the minimum of the values of $u$, corresponding to the
spatial volume. The emergence of new peaks in the noncommutative
wave packet may be interpreted as a representation of different
quantum states that may communicate with each other through
tunnelling. This  means that there are different possible universes
(states) from which our present universe could have evolved and
tunnelled in the past, from one universe (state) to another (see the
first reference in \cite{15}).
\begin{figure}
\begin{tabular}{ccc} \epsfig{figure=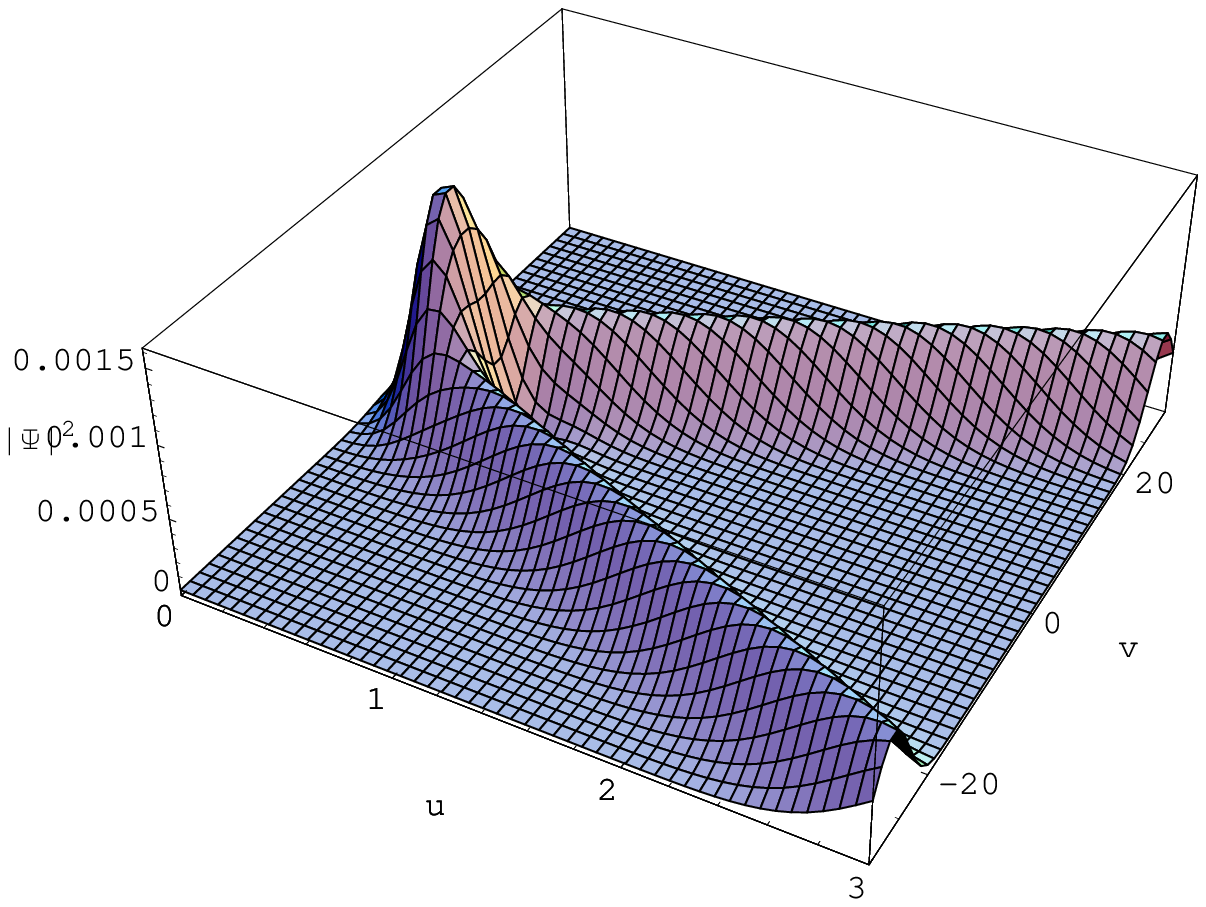,width=7cm}
\hspace{1cm} \epsfig{figure=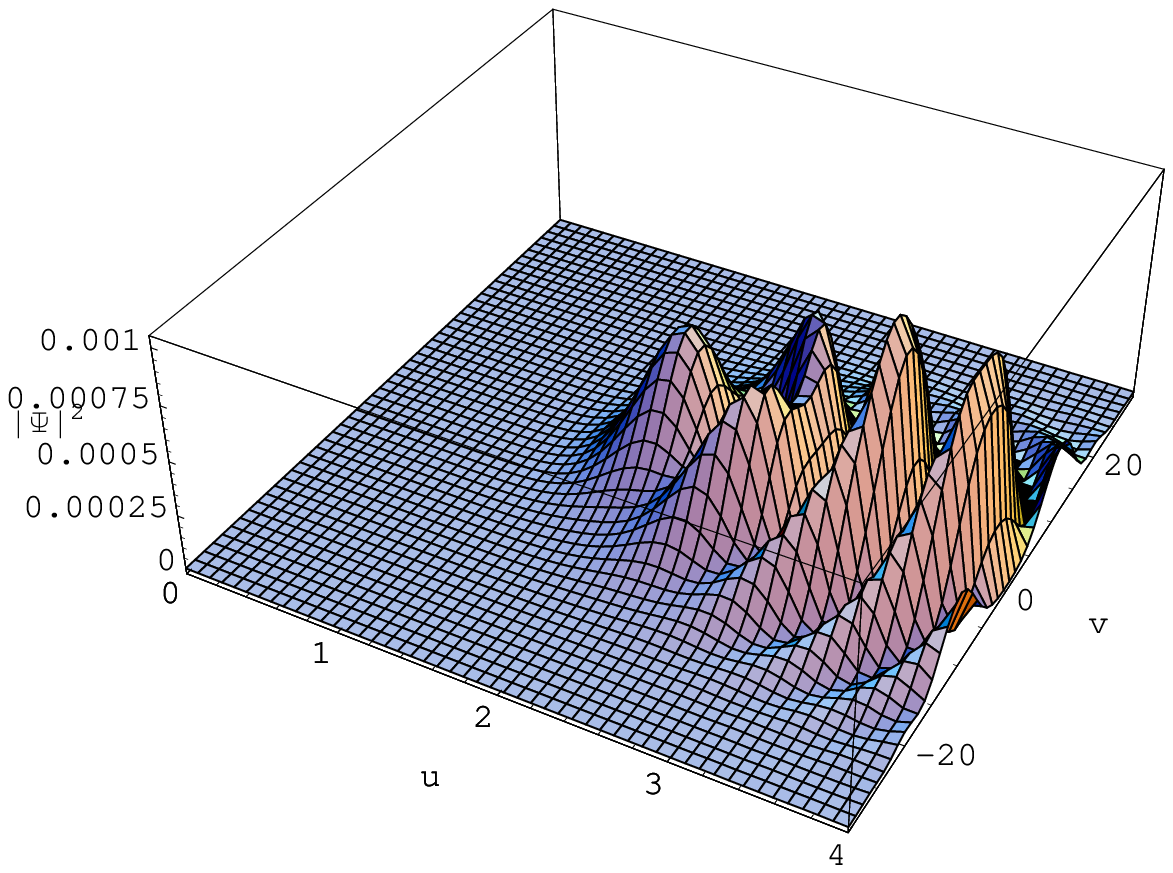,width=7cm}
\end{tabular}
\caption{\footnotesize The figure on the left shows the square of
the commutative wavefunction while the figure on the right, the
square of the noncommutative wavefunction. We choose $w=0$ for
simplicity.} \label{fig}
\end{figure}
\section{Symmetries of commutative and noncommutative Bianchi
class A spacetimes} In this section we investigate the Noether
Symmetries of the Bianchi class A models and study the effects of
noncommutativity on these symmetries. Following \cite{23} and
\cite{24}, we define the Noether symmetry of the spacetime as a
vector field $X$ on the tangent space of the phase-space-like
$(u,v,w,\dot{u},\dot{v},\dot{w})$ through
\begin{equation}\label{a}
X=\alpha \frac{\partial}{\partial u}+\beta \frac{\partial}{\partial
v}+\gamma \frac{\partial}{\partial w}+\frac{d
\alpha}{dt}\frac{\partial}{\partial \dot{u}}+\frac{d
\beta}{dt}\frac{\partial}{\partial \dot{v}}+\frac{d
\gamma}{dt}\frac{\partial}{\partial \dot{w}},
\end{equation} such that the Lie derivative of the Lagrangian with
respect to this vector field vanishes
\begin{equation}\label{b}
L_X{\cal L}=0.\end{equation} In (\ref{a}), $(\alpha,\beta,\gamma)$
are linear functions\footnote{One may, of course, seek more
complicated symmetries. However, we have chosen the linear case for
simplicity.} of $(u,v,w)$ and $\frac{d}{dt}$ represents the Lie
derivative along the dynamical vector field, that is
\begin{equation}\label{c}
\frac{d}{dt}=\dot{u}\frac{\partial}{\partial
u}+\dot{v}\frac{\partial}{\partial
v}+\dot{w}\frac{\partial}{\partial w}.
\end{equation}
It is easy to find the constants of motion corresponding to such
symmetry. Indeed, equation (\ref{b}) can be rewritten as
\begin{equation}\label{d}
L_X{\cal L}=\left(\alpha \frac{\partial {\cal L}}{\partial
u}+\frac{d \alpha}{dt}\frac{\partial {\cal L}}{\partial
\dot{u}}\right)+\left(\beta \frac{\partial {\cal L}}{\partial
v}+\frac{d \beta}{dt}\frac{\partial {\cal L}}{\partial
\dot{v}}\right)+\left(\gamma \frac{\partial {\cal L}}{\partial
w}+\frac{d \gamma}{dt}\frac{\partial {\cal L}}{\partial
\dot{w}}\right)=0.
\end{equation}
Noting that $\frac{\partial {\cal L}}{\partial q}=\frac{dp_q}{dt}$,
we have
\begin{equation}\label{e}
\left(\alpha
\frac{dp_u}{dt}+\frac{d\alpha}{dt}p_u\right)+\left(\beta
\frac{dp_v}{dt}+\frac{d\beta}{dt}p_v\right)+\left(\gamma
\frac{dp_w}{dt}+\frac{d\gamma}{dt}p_w\right)=0,
\end{equation}
which yields
\begin{equation}\label{f}
\frac{d}{dt}\left(\alpha p_u+\beta p_v+\gamma p_w\right)=0.
\end{equation}
Thus the constants of motion are found as
\begin{equation}\label{g}
Q=\alpha p_u+\beta p_v+\gamma P_w.
\end{equation}
In order to obtain the functions $\alpha$, $\beta$ and $\gamma$ we
use equation (\ref{d}).  However, since in the noncommutative case
we use the Hamiltonian formalism, equation (\ref{e}) is better
suited in finding these coefficients which, equivalently, can be
written as
\begin{eqnarray}\label{h}
\alpha \{p_u,{\cal H}\}+\beta \{p_v,{\cal H}\}+\gamma \{p_w,{\cal
H}\}+\left[\frac{\partial \alpha}{\partial u}\{u,{\cal
H}\}+\frac{\partial \alpha}{\partial v}\{v,{\cal H}\}+\frac{\partial
\alpha}{\partial w}
\{w,{\cal H}\}\right]p_u+\nonumber\\
\left[\frac{\partial \beta}{\partial u}\{u,{\cal H}\}+\frac{\partial
\beta}{\partial v}\{v,{\cal H}\}+\frac{\partial \beta}{\partial w}
\{w,{\cal H}\}\right]p_v+ \left[\frac{\partial \gamma}{\partial
u}\{u,{\cal H}\}+\frac{\partial \gamma}{\partial v}\{v,{\cal
H}\}+\frac{\partial \gamma}{\partial w} \{w,{\cal H}\}\right]p_w=0.
\end{eqnarray}
In general, the expression above gives a quadratic polynomial in
terms of momenta with coefficients being partial derivatives of
$\alpha$, $\beta$ and $\gamma$ with respect to the configuration
variables $u$, $v$ and $w$. Thus, the expression is identically
equal to zero if and only if these coefficients are zero, leading to
a system of partial differential equations for $\alpha$, $\beta$ and
$\gamma$. In the following subsections we obtain such symmetries for
Bianchi class A models in the commutative and noncommutative cases.
\subsection{The Bianchi type I model}
This model is the simplest in that all the structure constants are
zero. The Hamiltonian in the gauge $N=e^{3u}$ is given by equation
(\ref{L}). Substituting the Poisson brackets (\ref{O})-(\ref{Q}) in
equation (\ref{h}), we get
\begin{eqnarray}\label{i}
-6\Lambda \alpha e^{6u}-\frac{1}{12}\frac{\partial \alpha}{\partial
u}p_u^2+\frac{1}{12}\frac{\partial \beta}{\partial
v}p_v^2+\frac{1}{12} \frac{\partial \gamma}{\partial w}p_w^2+
\frac{1}{12}\left(\frac{\partial \alpha}{\partial v}-\frac{\partial
\beta}{\partial u}\right)p_up_v+ \frac{1}{12}\left(\frac{\partial
\alpha}{\partial
w}-\frac{\partial \gamma}{\partial u}\right)p_up_w\nonumber\\
+\frac{1}{12}\left(\frac{\partial \beta}{\partial w}+\frac{\partial
\gamma}{\partial v}\right)p_vp_w=0,
\end{eqnarray}
which leads to the following system
\begin{eqnarray}\label{j}
\alpha=0,\hspace{.5cm}\frac{\partial \alpha}{\partial u}=
\frac{\partial \beta}{\partial v}=\frac{\partial \gamma}{\partial w}=0,\nonumber \\
\\
\frac{\partial \alpha}{\partial v}-\frac{\partial \beta}{\partial
u}= \frac{\partial \alpha}{\partial w}-\frac{\partial
\gamma}{\partial u}= \frac{\partial \beta}{\partial
w}+\frac{\partial \gamma}{\partial v}=0. \nonumber
\end{eqnarray}
The above system has the general solutions \cite{23,24}
\begin{equation}\label{l}
\alpha=0,\hspace{.5cm}\beta=aw+b,\hspace{.5cm}\gamma=-av+b',
\end{equation}
where $a$, $b$ and $b'$ are constants. Thus, according to (\ref{g})
the independent constants of motion are
\begin{equation}\label{m}
Q_1=p_v,\hspace{.5cm}Q_2=p_w,\hspace{.5cm}Q_3=wp_v-vp_w,
\end{equation}
which correspond to the following symmetries
\begin{equation}\label{n}
X_1=\frac{\partial}{\partial
v},\hspace{.5cm}X_2=\frac{\partial}{\partial
w},\hspace{.5cm}X_3=w\frac{\partial}{\partial
v}-v\frac{\partial}{\partial w}+\dot{w}\frac{\partial}{\partial
\dot{v}}-\dot{v}\frac{\partial}{\partial \dot{w}}.
\end{equation}
These symmetries satisfy the Lie algebra
\begin{equation}\label{o}
[X_1,X_2]=0,\hspace{.5cm}[X_2,X_3]=X_1,\hspace{.5cm}[X_3,X_1]=X_2.
\end{equation}
Now, let us see which of the above symmetries survive in the
noncommutative case. Here, the Hamiltonian is given by (\ref{AS})
and the required Poisson brackets are given by equations
(\ref{AT})-(\ref{AV}). Substituting these Poisson brackets into
relation (\ref{h}) gives
\begin{eqnarray}\label{p}
-6\Lambda \alpha
e^{6u-3\theta_1p_v+3\theta_3p_w}-\frac{1}{12}\frac{\partial
\alpha}{\partial u}p_u^2+\frac{1}{12}\frac{\partial \beta}{\partial
v}p_v^2+\frac{1}{12}
\frac{\partial \gamma}{\partial w}p_w^2+\nonumber \\
\frac{1}{12}\left(\frac{\partial \alpha}{\partial v}-\frac{\partial
\beta}{\partial u}\right)p_up_v+ \frac{1}{12}\left(\frac{\partial
\alpha}{\partial w}-\frac{\partial \gamma}{\partial u}\right)p_up_w
+\frac{1}{12}\left(\frac{\partial \beta}{\partial
w}+\frac{\partial \gamma}{\partial v}\right)p_vp_w+\nonumber\\
3\Lambda
e^{6u-3\theta_1p_v+3\theta_3p_w}\left[\left(\theta_1\frac{\partial
\alpha}{\partial v}-\theta_3\frac{\partial \alpha}{\partial w}
\right)p_u+\left(\theta_1\frac{\partial \beta}{\partial
v}-\theta_3\frac{\partial \beta}{\partial
w}\right)p_v+\left(\theta_1\frac{\partial \gamma} {\partial
v}-\theta_3\frac{\partial \gamma}{\partial w}\right)p_w\right]=0.
\end{eqnarray}
Thus, in addition to equations (\ref{j}) we have a number of extra
restrictions on $\alpha$, $\beta$ and $\gamma$ as
\begin{equation}\label{q}
\theta_1\frac{\partial \alpha}{\partial v}-\theta_3\frac{\partial
\alpha}{\partial w}=\theta_1\frac{\partial \beta}{\partial
v}-\theta_3\frac{\partial \beta}{\partial w}=\theta_1\frac{\partial
\gamma}{\partial v}-\theta_3\frac{\partial \gamma}{\partial w}=0,
\end{equation}
which should hold for all values of $\theta_1$ and $\theta_3$.
Imposing relations (\ref{q}) on the solutions (\ref{j}) results in
$\beta$ and $\gamma$ becoming constants. Therefore, in the
noncommutative case the only constants of motions are
\begin{equation}\label{r}
(Q_1)_{nc}=p_v,\hspace{.5cm}(Q_2)_{nc}=p_w,
\end{equation}
which correspond to the symmetries
\begin{equation}\label{s}
(X_1)_{nc}=\frac{\partial}{\partial
v},\hspace{.5cm}(X_2)_{nc}=\frac{\partial}{\partial
w},\hspace{.5cm}[X_1,X_2]=0.
\end{equation}
Consequently, the third symmetry is absent in the noncommutative
case. A quick look at the Hamiltonians (\ref{L}) and (\ref{AS})
shows that the variables $v$ and $w$ are cyclic and consequently
the corresponding  momenta are constants of motion. In addition,
Hamiltonian (\ref{L}) is rotationally invariant about the
$u$-axis, that is $\dot{v}^2+\dot{w}^2$. Thus, the angular
momentum about this axis is conserved. This symmetry is absent in
the noncommutative Hamiltonian (\ref{AS}).

At this point, it is appropriate to investigate the symmetries
when the cosmological constant is zero. In this case there is no
difference between commutative and noncommutative Bianchi type I
models. It is clear from (\ref{i}) that the symmetries in this
case are obtained from the system (\ref{j}) without the constraint
$\alpha=0$ and that the solutions are given by
\begin{equation}\label{ac}
\alpha=av+bw+c,\hspace{.5cm}\beta=au+b'w+c',\hspace{.5cm}\gamma=bu-b'v+c'',
\end{equation}
with $a$, $b$, $b'$, $c$, $c'$ and $c''$ being constants. After a
simple calculation we find the independent constants of motion,
given by
\begin{equation}\label{ad}
\begin{array}{c}
Q_1=p_u,\hspace{.5cm}Q_2=p_v,\hspace{.5cm}Q_3=p_w,\vspace{.5cm} \\
Q_4=wp_v-vp_w,\hspace{.5cm}Q_5=wp_u+up_w,\hspace{.5cm}Q_6=vp_u+up_v, \\
\end{array}
\end{equation}which correspond to the following symmetries
\begin{eqnarray}\label{ae}
X_1&=&\frac{\partial}{\partial
u},\hspace{.5cm}X_2=\frac{\partial}{\partial v},
\hspace{.5cm}X_3=\frac{\partial}{\partial w},\nonumber \\
X_4&=&w\frac{\partial}{\partial v}-v\frac{\partial}{\partial
w}+\dot{w}\frac{\partial}{\partial
\dot{v}}-\dot{v}\frac{\partial}{\partial \dot{w}}
\vspace{.5cm} \nonumber\\
\\
X_5&=&w\frac{\partial}{\partial u}+u\frac{\partial}{\partial
w}+\dot{w}\frac{\partial}{\partial
\dot{u}}+\dot{u}\frac{\partial}{\partial \dot{w}}
\nonumber \\
X_6&=&v\frac{\partial}{\partial u}+u\frac{\partial}{\partial
v}+\dot{v}\frac{\partial}{\partial
\dot{u}}+\dot{u}\frac{\partial}{\partial \dot{v}}\nonumber,
\end{eqnarray}
satisfying the following Lie algebra
\begin{eqnarray}\label{af}
[X_1,X_2]&=&0,\hspace{.5cm}[X_1,X_3]=0,\hspace{.5cm}[X_1,X_4]=0,\nonumber\\
\left[X_1,X_5\right]&=&X_3,\hspace{.5cm}[X_1,X_6]=X_2,\hspace{.5cm}[X_2,X_3]=0,\nonumber \\
\left[X_2,X_4\right]&=&-X_3,\hspace{.5cm}[X_2,X_5]=0,\hspace{.5cm}[X_2,X_6]=X_1,\\
\left[X_3,X_4\right]&=&X_2,\hspace{.5cm}[X_3,X_5]=X_1,\hspace{.5cm}[X_3,X_6]=0,\nonumber \\
\left[X_4,X_5\right]&=&-X_6,\hspace{.5cm}[X_4,X_6]=X_5,
\hspace{.5cm}[X_5,X_6]=X_4.\nonumber
\end{eqnarray}
\subsection{The Bianchi type II model}
This model is characterized by the 1-forms
\begin{equation}\label{t}
\omega^1=dx-zdy,\hspace{.5cm}\omega^2=dy,\hspace{.5cm}\omega^3=dz.
\end{equation}
Therefore, the non-vanishing structure constants may be obtained
from equation (\ref{C})
\begin{equation}\label{u}
C^1_{23}=-C^1_{32}=1,
\end{equation}
which, upon substitution into relation (\ref{H}), yields the scalar
curvature of the corresponding 3-geometry as
\begin{equation}\label{v}
R=2e^{-2u+4v+4\sqrt{3}w}.
\end{equation}
Thus, the Hamiltonian for the model can be written from equation
(\ref{K}) with the result (again we use the gauge $N=e^{3u}$)
\begin{equation}\label{w}
{\cal H}=\frac{1}{24}\left(-p_u^2+p_v^2+p_w^2\right)+\Lambda
e^{6u}-2e^{4(u+v+\sqrt{3}w)}.
\end{equation}
In order to obtain the symmetries, we have to solve equation
(\ref{h}) for $\alpha$, $\beta$ and $\gamma$. The required Poisson
brackets are
\begin{eqnarray}\label{x}
\{p_u,{\cal H}\}&=&-\left[6\Lambda e^{6u}-8e^{4(u+v+\sqrt{3}w)}\right],\nonumber \\
\{p_v,{\cal H}\}&=&8e^{4(u+v+\sqrt{3}w)},\hspace{.5cm}\{p_w,{\cal H}\}=8\sqrt{3}e^{4(u+v+\sqrt{3}w)}, \\
\{u,{\cal H}\}&=&-\frac{1}{12}p_u,\hspace{.5cm}\{v,{\cal
H}\}=\frac{1}{12}p_v,\hspace{.5cm}\{w,{\cal
H}\}=\frac{1}{12}p_w.\nonumber
\end{eqnarray}
These Poisson brackets make equation (\ref{h}) read
\begin{eqnarray}\label{y}
-6\alpha \Lambda
e^{6u}+8(\alpha+\beta+\sqrt{3}\gamma)e^{4(u+v+\sqrt{3}w)}
-\frac{1}{12}\frac{\partial \alpha}{\partial
u}p_u^2+\frac{1}{12}\frac{\partial \beta}{\partial v}p_v^2
+\frac{1}{12}\frac{\partial \gamma}{\partial w}p_w^2+\nonumber\\
\frac{1}{12}\left(\frac{\partial \alpha}{\partial v}-\frac{\partial
\beta}{\partial u}\right)p_up_v +\frac{1}{12}\left(\frac{\partial
\alpha}{\partial w}-\frac{\partial \gamma}{\partial u}\right)p_up_w
+\frac{1}{12}\left(\frac{\partial \beta}{\partial w}+\frac{\partial
\gamma}{\partial v}\right)p_vp_w=0,
\end{eqnarray}
which, in turn, results in the following system
\begin{eqnarray}\label{z}
\alpha=0,\hspace{.5cm}\alpha+\beta+\sqrt{3}\gamma=0,\hspace{.5cm}\frac{\partial
\alpha}{\partial u}=\frac{\partial \beta}
{\partial v}=\frac{\partial \gamma}{\partial w}=0,\nonumber \\
\\
\frac{\partial \alpha}{\partial v}-\frac{\partial \beta}{\partial
u}=0,\hspace{.5cm} \frac{\partial \alpha}{\partial w}-\frac{\partial
\gamma}{\partial u}=0,\hspace{.5cm} \frac{\partial \beta}{\partial
w}+\frac{\partial \gamma}{\partial v}=0.\nonumber
\end{eqnarray}
The solutions of this system are similar to the solutions of the
system (\ref{j}) with the additional constraint
$\alpha+\beta+\sqrt{3}\gamma=0$. Therefore, we obtain $\alpha=0$,
$\beta$ and $\gamma=\mbox{const.}$ as solutions, \textit{i.e.} we
have $\alpha=0$, $\beta=-\sqrt{3}$ and $\gamma=1$. Consequently,
there is only one symmetry
\begin{equation}\label{aa}
X=-\sqrt{3}\frac{\partial}{\partial v}+\frac{\partial}{\partial w},
\end{equation}
with the corresponding constant of motion
\begin{equation}\label{ab}
Q=-\sqrt{3}p_v+p_w.
\end{equation}
In the noncommutative case we have the system of equations
(\ref{z}) plus the additional equations in terms of the
noncommutativity parameters and partial derivative of $\alpha$,
$\beta$ and $\gamma$, similar to the last term in (\ref{p}). It is
easy to see that these additional equations do not have any effect
on the solutions of system (\ref{z}), and thus, also hold in the
noncommutative phase-space.

Now, let us consider this model with a zero cosmological constant.
In this case the restriction $\alpha=0$ is removed from the system
of equations (\ref{z}) and it may be shown that the general
solutions can be written as
\begin{equation}\label{ag}
\alpha=\sqrt{3}av-aw-c-\sqrt{3}c',\hspace{.5cm}\beta=\sqrt{3}au+aw+c,\hspace{.5cm}\gamma=-au-av+c',
\end{equation}
where $a$, $c$ and $c'$ are constant. Therefore, in this case we
have three independent constants of motion
\begin{eqnarray}\label{ah}
Q_1=-\sqrt{3}p_u+p_w,\hspace{.5cm}Q_2=-p_u+p_v,\nonumber\\
\\
Q_3=\left(\sqrt{3}v-w\right)p_u+\left(\sqrt{3}u+w\right)p_v-(u+v)p_w,
\nonumber
\end{eqnarray}
which correspond to the following symmetries
\begin{eqnarray}\label{ai}
X_1=-\sqrt{3}\frac{\partial}{\partial u}+\frac{\partial}{\partial
w},\hspace{.5cm}X_2=
-\frac{\partial}{\partial u}+\frac{\partial}{\partial v},\nonumber \\
X_3=\left(\sqrt{3}v-w\right)\frac{\partial}{\partial
u}+\left(\sqrt{3}u+w\right)\frac{\partial}{\partial
v}-(u+v)\frac{\partial}{\partial w} +\\
\left(\sqrt{3}\dot{v}-\dot{w}\right)\frac{\partial}{\partial
\dot{u}}+\left(\sqrt{3}\dot{u}+\dot{w}\right)\frac{\partial}{\partial
\dot{v}} -(\dot{u}+\dot{v})\frac{\partial}{\partial
\dot{w}},\nonumber
\end{eqnarray}
satisfying the Lie algebra given by
\begin{equation}\label{aj}
[X_1,X_2]=0,\hspace{.5cm}[X_1,X_3]=\sqrt{3}X_1-2X_2,\hspace{.5cm}[X_2,X_3]=-\sqrt{3}X_2.
\end{equation}
The noncommutative phase-space with zero cosmological constant has
the Hamiltonian
\begin{equation}\label{al}
{\cal
H}=\frac{1}{24}\left(-p_u^2+p_v^2+p_w^2\right)-2e^{4u-2\theta_1
p_v+2\theta_3p_w}e^{4v+2\theta_1p_u-2\theta_2p_w}e^{\sqrt{3}(4w-2\theta_3p_u+2\theta_2p_v)}.
\end{equation}
To construct equation (\ref{h}) we need the following  Poisson
brackets
\begin{eqnarray}\label{am}
\{p_u,{\cal H}\}=\{p_v,{\cal H}\}=\frac{1}{\sqrt{3}}\{p_w,{\cal H}\}=8e^Ae^Be^C,\nonumber \\
\hspace{-.8cm}\{u,{\cal H}\}=-\frac{1}{12}p_u-4\left(\theta_1-\sqrt{3}\theta_3\right)e^Ae^Be^C,\nonumber \\
\\
\hspace{-1.1cm}\{v,{\cal H}\}=\frac{1}{12}p_v+4\left(\theta_1-\sqrt{3}\theta_3\right)e^Ae^Be^C,\nonumber \\
\hspace{-1.3cm}\{w,{\cal
H}\}=\frac{1}{12}p_w-4\left(\theta_3-\theta_2\right)e^Ae^Be^C,\nonumber
\end{eqnarray}
where $A=4u-2\theta_1p_v+2\theta_3p_w$,
$B=4v+2\theta_1p_u-2\theta_2p_w$ and
$C=\sqrt{3}(4w-2\theta_3p_u+2\theta_2p_v)$. Substitution of these
relations into equation (\ref{h}) shows that solutions (\ref{ag})
should satisfy the following additional constraints
\begin{eqnarray}\label{amm}
\left(\theta_1-\sqrt{3}\theta_3\right)\frac{\partial
\alpha}{\partial v}-(\theta_3-\theta_2)\frac{\partial
\alpha}{\partial w}=0,
\nonumber \\
\left(\theta_1-\sqrt{3}\theta_3\right)\frac{\partial \beta}{\partial
u}+(\theta_3-\theta_2)\frac{\partial \beta}{\partial w}=0, \\
\frac{\partial \gamma}{\partial u}-\frac{\partial \gamma}{\partial
v}=0. \nonumber
\end{eqnarray}
Imposing these conditions on solutions (\ref{ag}), we are led to
\begin{equation}\label{aam}
a\left(\sqrt{3}\theta_1-\theta_2-2\theta_3\right)=0.
\end{equation}
Now if $a=0$, \textit{i.e.} $\alpha$, $\beta$ and $\gamma$ are
constant, then solutions (\ref{ag}) automatically satisfy
conditions (\ref{amm}). Therefore, the noncommutative case has the
symmetries $X_1$ and $X_2$ only, given by (\ref{ai}). On the other
hand if $a\neq 0$, there is a special combination of $\theta$'s,
that is
\begin{equation}\label{am1}
\sqrt{3}\theta_1-\theta_2-2\theta_3=0,
\end{equation}
for which all the symmetries (\ref{ai}) in the commutative case
can also be recovered in the noncommutative case. This phenomena
can be understood by noting that in this case Hamiltonian
(\ref{al}) takes the form
\begin{equation}\label{am2}
{\cal
H}=\frac{1}{24}\left(-p_u^2+p_v^2+p_w^2\right)-2e^{2(\theta_1-\sqrt{3}\theta_3)
(2Q_2-\sqrt{3}Q_1)}e^{4(u+v+\sqrt{3}w)}.
\end{equation}
Since $Q_1$ and $Q_2$ are constants of motion the above Hamiltonian
differs from the Hamiltonian (\ref{w}) (for $\Lambda =0$) only by a
constant factor in the potential term. As it is clear from equation
(\ref{h}) which is the starting point in our procedure for finding
the symmetries, this constant factor does not have any effect on the
symmetries. Thus in the special case (\ref{am1}), Hamiltonians
(\ref{am2}) and (\ref{w}) have the same structure as far as the
symmetry considerations are concerned.
\subsection{The Bianchi types VI$_0$ and VII$_0$}
These types of Bianchi models are characterized by the 1-forms
\begin{equation}\label{an}
\omega^1=\cosh z dx \mp \sinh z dy,\hspace{.5cm}\omega^2=-\sinh z
dx+\cosh z dy,\hspace{.5cm}\omega^3=dz,
\end{equation}
with the corresponding structure constants
\begin{equation}\label{ao}
C^1_{23}=-C^1_{32}=\pm1,\hspace{.5cm}C^2_{31}=-C^2_{13}=-1,
\end{equation}
where the upper and lower signs denotes the Bianchi type VI$_0$ and
VII$_0$ respectively. The scalar curvature of the corresponding
3-geometry can be evaluated from equation (\ref{H}) with the result
\begin{equation}\label{ap}
R=4e^{u+4v}\left[\cosh (4\sqrt{3}w)\pm1 \right],
\end{equation}
leading to the Hamiltonian
\begin{equation}\label{ar}
{\cal H}=\frac{1}{24}\left(-p_u^2+p_v^2+p_w^2\right)+\Lambda
e^{6u}-4e^{4(u+v)}\left[\cosh (4\sqrt{3}w)\pm 1\right].
\end{equation}
To find the symmetries, we need the following Poisson brackets,
obtained from relation (\ref{h})
\begin{eqnarray}\label{as}
\{p_u,{\cal H}\}&=&-6\Lambda e^{6u}+16e^{4(u+v)}\left[\cosh (4\sqrt{3}w)\pm 1\right],\nonumber \\
\hspace{-1.8cm}\{p_v,{\cal H}\}&=& 16e^{4(u+v)}\left[\cosh (4\sqrt{3}w)\pm 1\right],\nonumber\\
\\
\hspace{-2.3cm}\{p_w,{\cal H}\}&=&16\sqrt{3}e^{4(u+v)}\sinh (4\sqrt{3}w),\nonumber \\
\hspace{.8cm}\{u,{\cal
H}\}&=&-\frac{1}{12}p_u,\hspace{.5cm}\{v,{\cal
H}\}=\frac{1}{12}p_v,\hspace{.5cm} \{w,{\cal
H}\}=\frac{1}{12}p_w.\nonumber
\end{eqnarray}
Substitution of the above Poisson brackets into equation (\ref{h})
leads to the following equation
\begin{eqnarray}\label{at}
-6\alpha \Lambda e^{6u}+16(\alpha+\beta)e^{4(u+v)}\left[\cosh (4\sqrt{3}w)\pm 1\right]+\nonumber \\
16\sqrt{3}\gamma e^{4(u+v)}\sinh
(4\sqrt{3}w)-\frac{1}{12}\frac{\partial \alpha}{\partial u}p_u^2+
\frac{1}{12}\frac{\partial \beta}{\partial v}p_v^2+\frac{1}{12}\frac{\partial \gamma}{\partial w}p_w^2+\nonumber\\
\frac{1}{12}\left(\frac{\partial \alpha}{\partial v}-\frac{\partial
\beta}{\partial u}\right)p_up_v+ \frac{1}{12}\left(\frac{\partial
\alpha}{\partial w}-\frac{\partial \gamma}{\partial u}\right)p_up_w+
\frac{1}{12}\left(\frac{\partial \beta}{\partial w}+\frac{\partial
\gamma}{\partial v}\right)p_vp_w=0,
\end{eqnarray}
which for a non-zero cosmological constant immediately yields
$\alpha=\beta=\gamma=0$, \textit{i.e.} Bianchi types VI$_0$ and
VII$_0$ universes with non-zero cosmological constants have no
symmetries and, therefore, no constants of motion. It is clear
that this also holds when we consider such spacetimes in a
noncommutative scenario.

Let us now assume that the cosmological constant in these models is
equal to zero. In this case, from equation (\ref{at}) we have
\begin{eqnarray}\label{au}
\gamma=0,\hspace{.5cm}\alpha+\beta=0,\hspace{.5cm}\frac{\partial
\alpha}{\partial u}=
\frac{\partial \beta}{\partial v}=\frac{\partial \gamma}{\partial w} &=&0,\nonumber \\
\\
\frac{\partial \alpha}{\partial v}-\frac{\partial \beta}{\partial
u}=0,\hspace{.5cm} \frac{\partial \alpha}{\partial w}-\frac{\partial
\gamma}{\partial u}=0,\hspace{.5cm} \frac{\partial \beta}{\partial
w}+\frac{\partial \gamma}{\partial v}&=&0,\nonumber
\end{eqnarray}
which admit the solution $\alpha=-\beta=\mbox{const.}$ Therefore,
the only symmetry and its corresponding constant of motion are
obtained as
\begin{equation}\label{av}
X=\frac{\partial}{\partial u}-\frac{\partial}{\partial
v},\hspace{.5cm}Q=p_u-p_v.
\end{equation}
The Hamiltonian in the noncommutative model for $\Lambda=0$ is
\begin{eqnarray}\label{ax}
{\cal
H}&=&\frac{1}{24}\left(-p_{u}^2+p_{v}^2+p_{w}^2\right)\nonumber\\
&-&4e^{(4u-2\theta_1 p_v+2\theta_3 p_w)}e^{(4v+2\theta_1
p_u-2\theta_2 p_w)} \left[\cosh (4\sqrt{3}w-2\sqrt{3}\theta_3
p_u+2\sqrt{3}\theta_2 p_v)\pm 1\right],
\end{eqnarray}
yielding the following Poisson brackets
\begin{eqnarray}\label{ay}
\{p_u,{\cal H}\}&=&\{p_v,{\cal H}\}=16e^{A+B}\left(\cosh C\pm
1\right),\hspace{.5cm}\{p_w,{\cal H}\}=16\sqrt{3}\sinh C,
\nonumber \\
\hspace{-1.3cm}\{u,{\cal H}\}&=&-\frac{1}{12}p_u-8\theta_1
e^{A+B}\left(\cosh C \pm 1\right)+8\theta_3\sqrt{3}e^{A+B}\sinh C,
\nonumber\\
\\
\hspace{-1.6cm}\{v,{\cal H}\}&=&\frac{1}{12}p_v+8\theta_1
e^{A+B}\left(\cosh C \pm 1\right)-8\theta_2\sqrt{3}e^{A+B}\sinh C,
\nonumber \\
\hspace{-4cm}\{w,{\cal H}\}&=&\frac{1}{12}p_w-8(\theta_3-\theta_2)
e^{A+B}\left(\cosh C \pm 1\right).\nonumber
\end{eqnarray}
Using these relations in equation (\ref{h}), we conclude that the
solutions of the system (\ref{au}) should also satisfy the following
additional conditions
\begin{eqnarray}\label{az}
\frac{\partial \alpha}{\partial v}-\frac{\partial \alpha}{\partial
u}=0,\hspace{.5cm} \theta_3\frac{\partial \alpha}{\partial
u}-\theta_2\frac{\partial \alpha}{\partial
v}=0,\hspace{.5cm}(\theta_3-\theta_2)
\frac{\partial \alpha}{\partial w}=0,\nonumber \\
\\
\frac{\partial \beta}{\partial v}-\frac{\partial \beta}{\partial
u}=0,\hspace{.5cm} \theta_3\frac{\partial \beta}{\partial
u}-\theta_2\frac{\partial \beta}{\partial
v}=0,\hspace{.5cm}(\theta_3-\theta_2) \frac{\partial \beta}{\partial
w}=0. \nonumber
\end{eqnarray}
If $\theta_2\neq \theta_3$, these equations have the previous
solutions $\alpha=-\beta=\mbox{const.}$, and thus the spacetime
has the same symmetry as given by (\ref{av}). However, in the case
where $\theta_2=\theta_3$, the above system admits the solutions
\begin{equation}\label{az1}
\alpha=-\beta=aw+b,
\end{equation}
with constants $a$ and $b$. The independent constants of motions are
then given by
\begin{equation}\label{az2}
Q_1=p_u-p_v,\hspace{.5cm}Q_2=w(p_u-p_v),
\end{equation}
corresponding to the following symmetries
\begin{equation}\label{az3}
X_1=\frac{\partial}{\partial u}-\frac{\partial}{\partial
v},\hspace{.5cm}X_2=w\left(\frac{\partial}{\partial
u}-\frac{\partial}{\partial
v}\right)+\dot{w}\left(\frac{\partial}{\partial
\dot{u}}-\frac{\partial}{\partial
\dot{v}}\right),\hspace{.5cm}[X_1,X_2]=0.
\end{equation}
Here, we have an additional constant of motion relative to the
commutative case. This is notable in Hamiltonian (\ref{ax}) in that
when $\theta_2=\theta_3$, it takes the form
\begin{equation}\label{az4}
{\cal
H}=\frac{1}{24}\left(-p_u^2+p_v^2+p_w^2\right)-4e^{2\theta_1Q_1}e^{4(u+v)}\left[\cosh
(4\sqrt{3}w-2\theta_2\sqrt{3}Q_1)\pm 1\right].
\end{equation}
The above Hamiltonian differs from  Hamiltonian (\ref{ar}) (for
$\Lambda=0$) not only by a constant factor in the potential term
(which has no effect on the number of symmetries like the Bianchi
type II case) but also in the argument of the ``$\cosh$'' term. This
argument is modified by a constant shift which results in a new
symmetry (compare (\ref{az2}) with (\ref{av})).
\subsection{The Bianchi types VIII and IX}
In the Bianchi type VIII model the invariant 1-forms are
\begin{eqnarray}\label{az5}
\omega^1&=&\cosh y \cos z dx-\sin z dy,\nonumber \\
\omega^2&=&\cosh y \sin z dx+\cos z dy,\vspace{.5cm}\\
\omega^3&=&\sinh y dx+dz, \nonumber
\end{eqnarray}
and thus the corresponding non-vanishing structure constants can be
evaluated from (\ref{C}) with the result
\begin{equation}\label{az6}
C^1_{23}=-C^1_{32}=-1,\hspace{.5cm}C^2_{31}=-C^2_{13}=-1,\hspace{.5cm}C^3_{12}=-C^3_{21}=1.
\end{equation}
The scalar curvature of the spatial hypersurface can be obtained by
substituting the above structure constants in (\ref{H}), yielding
\begin{equation}\label{bz1}
R=-4e^{-2u+4v}\cosh (4\sqrt{3}w)-2e^{-2u-8v}+4e^{-2u-2v}\cosh
(2\sqrt{3}w)-2e^{-2u+4v}.
\end{equation}
Therefore, we can write the Hamiltonian using (\ref{L})
\begin{equation}\label{bz2}
{\cal H}=\frac{1}{24}\left(-p_u^2+p_v^2+p_w^2\right)+\Lambda
e^{6u}+4e^{4(u+v)}\cosh (4\sqrt{3}w)-4e^{2(2u-v)}\cosh
(2\sqrt{3}w)+2e^{4(u-2v)}+2e^{4(u+v)}.
\end{equation}
As it is clear from the above form, the spacetime has no symmetry.
Indeed, evaluating the required Poisson brackets and constructing
equation (\ref{h}), we see that this system admits only the
trivial solution $\alpha=\beta=\gamma=0$, \textit{i.e.} there are
no symmetries neither in the commutative nor noncommutative
models.

Our final goal is to address the same issues in the Bianchi type IX
model, characterized by the structure constants
\begin{equation}\label{bz3}
C^i_{jk}=\varepsilon _{ijk}.
\end{equation}
Calculations similar to the previous models lead to the following
Hamiltonian
\begin{equation}\label{bz4} {\cal
H}=\frac{1}{24}\left(-p_u^2+p_v^2+p_w^2\right)+\Lambda e^{6u}
+e^{4(u-2v)}-4e^{2(2u-v)}\cosh (2\sqrt{3}w)+2e^{4(u+v)}\left[\cosh
(4\sqrt{3}w)-1\right].
\end{equation}
Again, a glance at the equations resulting from the above
Hamiltonian shows that this spacetime has no symmetry, neither in
the commutative nor in its noncommutative version, even if the
cosmological constant is equal to zero.
\section{Conclusions}
In this paper we have studied the Bianchi type I classical and
quantum cosmology in both commutative and noncommutative
scenarios. We have obtained exact solutions of the vacuum
gravitational field equations in three cases when the cosmological
constant is positive, negative or zero. The corresponding
classical cosmology shows a singularity at $t\rightarrow \infty$,
together with increasing anisotropic parameters $v$ and $w$. We
have seen that noncommutativity can not exclude such behavior from
the solutions. The corresponding quantum cosmology of the model in
both commutative and noncommutative cases is obtained from the
exact solutions of the WD equation, showing how such solutions
differ.  

We have also studied the Noether symmetries of the Bianchi class A
models and investigated their noncommutative counterparts. In
Bianchi type I, we have shown that for a non-zero cosmological
constant, there are three Noether symmetries, two of which retain
their character in the noncommutative case. For a zero
cosmological constant, there is no difference between commutative
and noncommutative cases and the number of symmetries is six. The
constants of motion in this case are the momenta in $u$, $v$ and
$w$ directions and  their corresponding angular momenta. However,
we note that because of the wrong sign (minus) of $p_u$ in the
Hamiltonians, this dynamical variable behaves like a ghost field
and thus the angular momenta-like constants $Q_5$ and $Q_6$ do not
have the usual sign. The Bianchi type II model with a non-zero
cosmological constant in the commutative and noncommutative cases
has only one symmetry, which is a Killing vector field. If the
cosmological constant is zero, this spacetime has three symmetries
when studied in its noncommutative form. We have shown that when
the phase-space has a noncommutative structure, one of its
symmetries is removed except when the noncommutative parameters
satisfy a certain relation. With this relation between the
noncommutative parameters, although the Hamiltonians are different
in the commutative and noncommutative cases, they show the same
structure and thus exhibit the same symmetries.

The Bianchi types VI$_0$ and VII$_0$ with non-zero cosmological
constants have no symmetries neither in the commutative nor
noncommutative versions. But in the case of a zero cosmological
constant, these spacetimes exhibit one symmetry if their
phase-space is commutative. This symmetry remains unchanged in the
noncommutative phase-space as well. It is interesting to note that
if the noncommutative parameters $\theta_2$ and $\theta_3$ are
equal, these models have an additional symmetry. In this case the
corresponding Hamiltonians of the commutative and noncommutative
phase-space have the same structure but differ in a shift term in
the variable $w$. In the last section of the paper we have dealt
with the Bianchi type VIII and IX models and shown that they have
no symmetries at all.\vspace{6mm}\noindent\\
{\bf Acknowledgement}\vspace{2mm}\noindent\\
The authors would like to thank S. Jalalzadeh for useful comments
and the anonymous referees for enlightening suggestions.

\end{document}